\documentclass[aps,twocolumn,superscriptaddress]{revtex4}
\usepackage{latexsym}
\usepackage{graphicx}
\usepackage{amsmath}

\begin{document}

\title{Driving force of the orbital-relevant electronic nematicity in Fe-based superconductors}

\author{Tao Li}
\affiliation{ Department of Physics, Renmin University of China, Beijing 100872, P. R. China}

\author{Yuehua Su}
\email{suyh@ytu.edu.cn}
\affiliation{ Department of Physics, Yantai University, Yantai 264005, P. R. China}






\begin{abstract}
The electronic nematic responses in Fe-based superconductors have been observed ubiquitously 
in various experimental probes. One novel nematic character is the d-wave {\it bond} 
orbital-relevant nematic charge order which was firstly proposed by symmetry analysis and 
then conformed by angle-resolved photoemission spectroscopy. In this paper, we present a mechanism that 
the driving force of the orbital-relevant nematic charge order is the reduction of the large Hubbard 
energy in the particle-hole charge channel by virtual hopping processes. This is one scenario 
from strong-coupling consideration. The same virtual hopping processes can lead to a 
super-exchange interaction for the spin magnetic order in the particle-hole spin channel and 
a pairing interaction for the superconducting order in the particle-particle channel. Thus the 
electronic nematic order, the spin magnetic order and the pairing superconducting order are 
intrinsically entangled and they can all stem from the same microscopic virtual hopping processes 
in reduction of the Hubbard energy. The electronic nematicity, the spin magnetism and the pairing 
superconductivity in unconventional superconductors are proposed to be unified within this mechanism.     
\end{abstract}



\maketitle


The electronic nematic state is one universal electronic state with spontaneous rotational symmetry 
breaking~\cite{Fradkin}. It shows novel unconventional physics~\cite{FradkinNematicNonFL2001,
LawlerNonFLnematic2006,DellAnnaNematicNonFL2006,WuCJNematicFL2007,WatanabeVishwanath}. 
Recently, the electronic nematic state has been observed in Fe-based superconductors (FeSCs)
with ubiquitous nematic responses in various experimental probes~\cite{FisherShen,BohmerMeingast2016,
GallaisPaul2016,SuPLA2016}. One special character of the electronic nematicity in FeSCs is the 
unusual momentum-dependent electronic band deformation with large value at $(\pi,0)/(0,\pi)$ and 
small value at $\Gamma$ point as observed in ARPES~\cite{ZXShen,YiShen2011,YiShen2014,DingH2015,LuDH2015}. 
This special nematic character rules out the on-site orbital order of the different occupation of 
the $d_{xz}$ and $d_{yz}$ electrons, which is confirmed consistently by the Raman scattering in $B_{1g}$ 
channel~\cite{GallaisPaul2016,KretzschmarHackl2016} and the other ARPES data~\cite{SuzukiARPESPRB2015,
ColdeaWatson2017}. This novel nematic character has been attributed to a d-wave bond orbital-relevant 
nematic charge order which was firstly proposed by our symmetry analysis~\cite{SuLi2013,SuLi2014JPCM}. 
The orbital-relevant nematic charge order in FeSCs can enhance the magnetic condensation energy~\cite{SuLi2014JPCM}. 
It can also lead to the static~\cite{SuLi2013} and the dynamical~\cite{SuPLA2016} magnetic nematic 
responses which have been observed in the magnetic torque~\cite{KasahasiMatsuda} and the neutron 
scattering spectra~\cite{LuDaiScience2014}. Thus the orbital-relevant electronic nematicity 
plays important roles in the low-energy physics of FeSCs.

The driving force of the ubiquitous nematicity in FeSCs is still elusive.  The driving nematic 
order may be an Ising spin nematic order~\cite{FangHu2008,CKXu,HuXu}, an itinerant magnetic nematic 
order~\cite{Fernandes2012,Fernandes2014}, an orbital-relevant nematic charge order~\cite{SuLi2013,SuLi2014JPCM}, 
or a Pomeranchuk nematic order of Fermi surface instability~\cite{Zhai2009,ChubukovPRX2016,ChubukovPRB942016,
ChubukovPRL2017,ChubukovPRB952017}. Other possible driving mechanisms of the electronic 
nematicity are also proposed, such as to reduce the Huud's coupling energy~\cite{Fanfarillo2016} 
or to reduce the nearest-neighbour Hubbard energy~\cite{HuCPL2015,WangZQPRB2016,HuArXiv2016}. 

In this paper, we will study the driving force of the orbital-relevant electronic nematicity 
in FeSCs. Our starting point is the multi-orbital Hubbard model with relatively large Hubbard 
interactions to account for the relatively strong correlation of the 3d electrons in FeSCs. 
This is one scenario for FeSCs from strong-coupling consideration. The one-orbital Hubbard model
with strong Hubbard interaction has been extensively studied in understanding the unconventional 
superconductivity and the novel normal states of cuprate superconductors in the last three 
decades~\cite{PALeeRMP2006}. Within this scenario, the bond orbital-relevant nematic charge order 
can arise from the virtual hopping processes in reduction of the large Hubbard energy in the 
particle-hole charge channel. The same virtual hopping processes can lead to a super-exchange 
interaction for the spin magnetic order in the particle-hole spin channel and an effective pairing 
interaction for the superconducting order in the particle-particle channel. Thus the electronic 
nematic order, the spin magnetic order and the pairing superconducting order are unified to stem 
from the same microscopic virtual hopping processes. We also propose that the electronic nematicity, 
the spin magnetism and the pairing superconductivity in unconventional superconductors come from 
the same mechanism of reduction of the Hubbard energy by virtual hopping processes. 


Before a further study on the driving force of the orbital-relevant electronic nematicity in 
FeSCs, we firstly summarize the most possibly relevant nematic operators from the 
Landau's principle of symmetry breaking~\cite{LandauBook}. 
With a structural phase transition from a tetragonal to an orthorhombic phase,
the Fe-site symmetry group is reduced from $D_{2d}$ to $D_2$~\cite{SuLi2013,SuLi2014JPCM}.  
Following the Landau's principle, the electronic nematicity can be described by introducing 
a symmetry breaking term $\delta H$ into the Hamiltonian $H_0$ which belongs to the 
identity representation of the $D_{2d}$. Now the Hamiltonian $H = H_0 + \delta H$ 
belongs to the identity representation of the $D_2$ but not that of the $D_{2d}$. 
$\delta H$ can be expanded by the irreducible representations of the 
$D_{2d}$~\cite{LandauBook,SuPLA2016}, i.e.,
\begin{equation}
\delta H = \sum_{j,n,\mu} h^{(j)}_{n,\mu} O^{(j)}_{n,\mu} , \label{eqn1}
\end{equation}
where $\{ O^{(j)}_{n,\mu}, \mu = 1, 2, \cdots, d_j\}$ form the basis functions of the 
$j$-th irreducible representation of the $D_{2d}$, $d_j$ is the corresponding dimension. 
$n$ denotes the different groups of the $j$-th representation.  $h$ can be regarded as 
the conjugate field of the nematic operator $O$. This expansion can be easily 
obtained by projection operations and the summation does not include the identity 
representation of the $D_{2d}$. It is shown that there is only one channel in the 
symmetry reduction from $D_{2d}$ to $D_2$, i.e., $\delta H$ belongs to the $B_1$ 
representation of the $D_{2d}$.

\begin{table}[h]
\caption{ Orbital-relevant nematic operators in the particle-hole charge channel 
with $\{d_{xz}, d_{yz}\}$ orbitals involved. The matrix functions $\Xi(\mathbf{k})$ 
in Eq. (\ref{eqn1}) are defined as $\Xi(\mathbf{k}) = \Lambda(\mathbf{k}) \Omega$,  
where the matrices $\Omega$ are defined by the Pauli matices $\tau_i$ in the orbital 
space of $\{d_{xz}, d_{yz}\}$. The representation (repst.) of $D_{2d}$ in the last 
column are shown as $G_1 \otimes G_2 = G_3$, where $G_1, G_2$ and $G_3$ are the 
symmetries of $\Lambda(\mathbf{k})$, $\Omega$ and $\Xi(\mathbf{k})$, respectively. 
nn -- nearest-neighbour, nnn -- next-nearest-neighbour. }\label{tab1}
\begin{ruledtabular}
\begin{tabular}{llll}
  & $\Lambda(\mathbf{k})$ & $\Omega$ & repst. of $D_{2d}$ \\
\hline
on-site & c=const.  & $\tau_3$ & $A_1 \otimes B_1 = B_1$ \\
nn-bond & $\cos k_x + \cos k_y $ & $\tau_3$  &  $A_1 \otimes B_1 = B_1$ \\
  & $\cos k_x - \cos k_y $ & $\tau_0$  &  $B_1 \otimes A_1 = B_1$ \\  
nnn-bond & $\cos k_x  \cos k_y $ & $\tau_3$  &  $A_1 \otimes B_1 = B_1$ \\
  & $\sin k_x  \sin k_y $ & i$\tau_2$  &  $B_2 \otimes A_2 = B_1$ 
\end{tabular}
\end{ruledtabular}
\end{table}

\begin{table}[h]
\caption{ Orbital-relevant nematic operators in the particle-hole charge channel with 
three 3d orbitals $\{d_{xz}, d_{yz}, d_{xy}\}$ near Fermi energy involved.  The pure 
$\{d_{xz}, d_{yz}\}$-relevant ones are shown in Table (\ref{tab1}). The representation
(repst.) of $D_{2d}$ in the last column are shown as $G_1 \otimes G_2 = G_3$ with
$G_1$ and $G_2$ the symmetries of the components of $O(\mathbf{k})$.  The first two 
rows are for nearest-neighbour bond and the third row is for next-nearest-neighbour 
bond.  } \label{tab2}
\begin{ruledtabular}
\begin{tabular}{cc}
  $O(\mathbf{k})$  & repst. of $D_{2d}$ \\
\hline
 $\left(\cos k_x - \cos k_y\right) d^{\dag}_{\mathbf{k},xy} d_{\mathbf{k},xy} $  
                            &  $B_1 \otimes A_1 = B_1$ \\
 $ \left( d^{\dag}_{\mathbf{k},xz}, d^{\dag}_{\mathbf{k},yz}\right) i\tau_2 
  \left(\begin{array}{l}
        \sin k_y \\
        \sin k_x
        \end{array}  
  \right) d_{\mathbf{k},xy} $ & $A_2 \otimes B_2 = B_1$ \\
 $ \left( d^{\dag}_{\mathbf{k},xz}, d^{\dag}_{\mathbf{k},yz}\right) i\tau_2 
  \left( \begin{array}{l}
        \cos k_x \sin k_y \\
        \sin k_x \cos k_y
        \end{array}  
  \right) d_{\mathbf{k},xy} $  & $A_2 \otimes B_2 = B_1$
\end{tabular}
\end{ruledtabular}
\end{table}

In Tables (\ref{tab1}) and (\ref{tab2}), we present the on-site, the nearest-neighbour bond 
and the next-nearest-neighbour bond orbital-relevant nematic operators in the particle-hole 
charge channel. They are defined as 
\begin{equation}
O(\mathbf{k}) = \sum_{ab\sigma} d^{\dag}_{\mathbf{k}a\sigma} \Xi_{ab}(\mathbf{k}) 
d_{\mathbf{k}b\sigma} , \label{eqn2}
\end{equation}
where $a,b$ are the orbital indices and $\sigma$ are the spin indices. Three orbitals near 
Fermi energy, $\{d_{xz}, d_{yz}, d_{xy}\}$, are considered. The lattice versions of the 
orbital-relevant nematic charge operators with five 3d orbitals have been provided in the 
articles~\cite{SuLi2013,SuLi2014JPCM}. The d-wave orbital-relevant nematic charge operator,
$O(\mathbf{k}) = \sum_{a\sigma} (\cos k_x - \cos k_y) d^{\dag}_{\mathbf{k}a\sigma}d_{\mathbf{k}a\sigma}$
with $\{d_{xz}, d_{yz}\}$ orbitals involved, has been introduced to account for the unusual 
electronic band deformation as observed in ARPES~\cite{SuLi2013,SuLi2014JPCM,HuCPL2015,WangZQPRB2016}. 
As a simple d-wave orbital-relevant nematic charge order cannot account for all the details 
of the band deformation, especially near $\Gamma$ point, here we present the most possibly 
orbital-relevant nematic operators for further investigation of the electronic band deformation. 
It should be noted that the combination of the nearest-neighbour extended s-wave and d-wave  
bond orbital-relevant nematic charge orders can lead to a band deformation with both $(\pi,0)/(0,\pi)$ 
and $\Gamma$ points involved. Moreover, since the next-nearest-neighbour hopping integrals of 
the 3d electrons in FeSCs are not ignorable in magnitude, the next-nearest-neighbour bond 
orbital-relevant nematic charge operators may also arise, with one example defined as
$O(\mathbf{k}) = \sum_{\sigma} (\cos k_x \cos k_y) \left( d^{\dag}_{\mathbf{k},xz,\sigma}
d_{\mathbf{k},xz,\sigma} - d^{\dag}_{\mathbf{k},yz,\sigma}d_{\mathbf{k},yz,\sigma}\right)$ 
which has also been proposed recently in study of the Huud's coupling in FeSCs~\cite{Fanfarillo2016}.    

The orbital-relevant nematic operators defined in Eq. (\ref{eqn2}) are the on-site or
the bond ones in the particle-hole charge channel. There are many available orbital-relevant 
nematic operators in other channels. One special example is the orbital-relevant nematic 
operator in the particle-hole spin channel defined by
\begin{equation}
\boldsymbol{S}_{ij} = \sum_{a\sigma_1\sigma_2} p_{a} d^{\dag}_{ja\sigma_1} 
\left(\frac{\boldsymbol{\sigma}}{2}\right)_{\sigma_1\sigma_2} d_{ia\sigma_2} , \label{eqn3}
\end{equation} 
where $p_{a}$ is an orbital-relevant phase factor and $\boldsymbol{\sigma}$ are the Pauli 
matrices. It should be noted that the nematic operators we have introduced in Eq. (\ref{eqn2}) 
and (\ref{eqn3}) are the extended orbital-relevant Pomeranchuk nematic operators in the 
respective charge and spin channels. The original Pomeranchuk nematic operators are defined 
in band models without explicit orbital character as 
\begin{equation*}
n_{\mathbf{k}} = \sum_{\sigma}\phi_{\mathbf{k}} d^{\dag}_{k\sigma} d_{k\sigma} , \quad  
\boldsymbol{S}_{\mathbf{k}} = \sum_{\sigma_1\sigma_2}\phi_{\mathbf{k}} d^{\dag}_{k\sigma_1} 
\left(\frac{\boldsymbol{\sigma}}{2}\right)_{\sigma_1\sigma_2} d_{k\sigma_2} , 
\end{equation*}
where $\phi_{\mathbf{k}}$ is a $\mathbf{k}$-dependent form factor~\cite{Pomeranchuk1959}. 
The recent functional and parquet renormalization group studies found a Pomeranchuk nematic order 
in FeSCs with d-wave Fermi surface deformation~\cite{Zhai2009,ChubukovPRX2016,ChubukovPRB942016,
ChubukovPRL2017,ChubukovPRB952017}. It should be noted that what exact form factor of the 
proposed Pomeranchuk nematic order can not be explicitly presented in the renormalization group 
formulations since these theoretical formulations are established only near Fermi pockets.    
 
Any other operator which follows Eq. (\ref{eqn1}) of the Landau's principle can be taken 
as nematic operator. Examples in higher-order charge or spin channels can be defined
such as 
\begin{eqnarray}
&&O_n = \sum_{i a} \left( n_{ia} n_{i+\delta_x a} - n_{ia} n_{i+\delta_y a} \right) ,\nonumber \\
&&O_s = \sum_{i a} \left( \boldsymbol{S}_{ia} \cdot \boldsymbol{S}_{i+\delta_x a} 
- \boldsymbol{S}_{ia} \cdot \boldsymbol{S}_{i+\delta_y a} \right) .\nonumber
\end{eqnarray}
$O_s$ is the Ising spin nematic operator introduced to account for the 
spin nematic responses~\cite{FangHu2008,CKXu,HuXu}.
The diverse nematic responses in different channels stem from the diverse nematic
operators which manifest themselves by a finite coupling to the driving
nematic order operator.


Now let us study the driving force of the orbital-relevant electronic nematicity in FeSCs. 
Our starting point is the Hubbard model with relatively large Hubbard interactions. 
This assumption is based upon the following experimental results: 
i) the linear-temperature dependent magnetic susceptibility shows strong local spin magnetic 
fluctuations~\cite{GMZhang}; ii) both the local magnetic moment and the ordered magnetic moment 
are large~\cite{KuWPRL2015}; iii) the electronic band structure near Fermi energy is largely 
renormalized~\cite{MaletzPRB2014,WatsonPRB2015}. These experimental results together with 
the theoretical studies~\cite{SiQMNaturePhys2009,YinKotliar2011} show obviously the relatively 
strong correlation of the 3d electrons in FeSCs. 
 
Let us firstly focus on the one-orbital Hubbard model with Hamiltonian 
$H=-\sum_{\langle ij \rangle \sigma} t d^{\dag}_{i\sigma} d_{j\sigma}
+ U\sum_{i} n_{i\uparrow} n_{i\downarrow}$, where the hopping integrals are only defined
on the nearest-neighbour bonds. In the large-$U$ limit near half-filling, the Hubbard model 
involves the Mott physics which are proposed to be essential in cuprate superconductors~\cite{PALeeRMP2006}. 
This is a strong-coupling scenario for cuprate superconductors. In the large-$U$ limit near 
half-filling, the virtual hopping processes between the paired singly-occupied sites lead 
to an effective Hamiltonian in second-order perturbations,
$H_{eff} = -V\sum_{\langle ij \rangle\sigma_i}  P d^{\dag}_{i\sigma_2} 
d_{j\sigma_2} d^{\dag}_{j\sigma_1} d_{i\sigma_1} P$, where $P$ is a Gutzwiller projector 
and $V=\frac{2t^2}{U}$. The relevant physics of this effective Hamiltonian are extensively 
studied for cuprate superconductors~\cite{PALeeRMP2006}. 
There are particle-hole charge physics involved in $H_{eff}$, which can be described by 
\begin{equation}
H_{\chi} = -V \sum_{\langle ij \rangle} \chi_{ij}^{\dag} \chi_{ij} , \label{eqn4.1}
\end{equation}
where a bond charge operator $\chi_{ij}$ is defined in the particle-hole charge channel as 
\begin{equation}
\chi_{ij} = \sum_{\sigma}d^{\dag}_{j\sigma} d_{i\sigma} . \label{eqn4.2}
\end{equation}
The projector $P$ is not explicitly shown in Eq. (\ref{eqn4.1}). The Pomeranchuk charge 
operator $n_{\mathbf{k}}$ is the momentum version of the bond charge operator $\chi_{ij}$ 
with a local form factor. The bond charge order parameters, such as a d-wave nematic charge 
order, an extended s-wave charge order, or a staggered flux order with special local form 
factor, have been proposed in the study of the $t-J$ model for cuprate superconductors
~\cite{YamaseKohnotJmodel2000,PALeeRMP2006}. 
The virtual hopping processes also involve the physics in the spin channel described by a 
super-exchange interaction   
\begin{equation}
H_{s} = 2V \sum_{\langle ij\rangle} \left( \boldsymbol{S}_i \cdot \boldsymbol{S}_j 
+\frac{1}{4} n_i n_j \right). \label{eqn5}
\end{equation} 
Obviously, an antiferromagnetic order is favoured at low temperature.  
The virtual hopping processes also involve the pairing physics in the particle-particle channel. 
They can be described by a new expression of $H_{eff}$ as 
\begin{eqnarray}
H_{\Delta} &=& V \sum_{\langle ij \rangle} (\Delta_{ij}^{(s,1)\dag} \Delta^{(s,1)}_{ij} 
+ \Delta_{ij}^{(s,-1)\dag} \Delta^{(s,-1)}_{ij}
\nonumber \\
&+& \frac{1}{2} \Delta_{ij}^{(s,0)\dag} \Delta^{(s,0)}_{ij} 
- \frac{1}{2} \Delta_{ij}^{(a)\dag} \Delta^{(a)}_{ij} ) , \label{eqn6}
\end{eqnarray} 
where the pairing operators are defined as $\Delta^{(s,1)}_{ij} = d_{j\uparrow} d_{i\uparrow}$, 
$\Delta^{(s,-1)}_{ij} = d_{j\downarrow} d_{i\downarrow}$, and 
$\Delta^{(s,0)/(a)}_{ij} = (d_{j\uparrow} d_{i\downarrow} \pm d_{j\downarrow} d_{i\uparrow})$.
The pairing Hamiltonian $H_{\Delta}$ shows that a stable singlet pairing state can 
come from the virtual hopping processes. 

This simple analysis on the one-orbital Hubbard model from strong-coupling consideration shows 
that the same virtual hopping processes in reduction of the on-site Hubbard energy involve various 
physics in different channels. Therefore the bond charge order, the spin magnetic order and the 
pairing superconducting order are unified to stem from the same microscopic virtual hopping processes 
but in different channels. This is our principal idea on the driving force of the bond nematic charge 
order in FeSCs and its intrinsic correlation with the spin magnetism and the pairing superconductivity.

The multi-orbital Hubbard model for FeSCs is simplified as 
$H=-\sum_{(ij) ab\sigma} t^{ab}_{ij} d^{\dag}_{ia\sigma} d_{jb\sigma}
+ U\sum_{i a} n_{i a\uparrow} n_{i a\downarrow}+ U\sum_{i,a\not= b} n_{ia} n_{ib}$,
where $(ij)$ include both the nearest-neighbour and the next-nearest-neighbour hoppings 
of the 3d electrons, and the on-site chemical potential terms are not explicitly shown. 
In the case with large $U$, the virtual hopping processes in second-order perturbations
with singly-occupied orbitals  involved can be described by an effective Hamiltonian 
$H^{\prime}_{eff} = - \sum_{ ( ij ) ab\sigma_i} V^{ab}_{ij} P 
d^{\dag}_{ia\sigma_2} d_{jb\sigma_2} d^{\dag}_{jb\sigma_1} d_{ia\sigma_1} P $, where 
$V^{ab}_{ij} = \frac{2\left( t^{ab}_{ij}\right)^2}{U}$ and $a, b$ are the indices of the 
singly-occupied orbitals. In the particle-hole charge channel, $H^{\prime}_{eff}$ can be 
rewritten as 
\begin{equation}
H^{\prime}_{\chi} = - \sum_{( ij ) ab} V^{ab}_{ij} 
\chi_{ij,ab}^{\dag} \chi_{ij,ab} ,  \label{eqn7.1}
\end{equation}
where the orbital-relevant bond charge operator is defined as 
$\chi_{ij,ab} = \sum_{\sigma} d^{\dag}_{jb\sigma} d_{ia\sigma}$.
Of all the singly-occupied orbitals involved in $H^{\prime}_{\chi}$, if two orbitals are
degenerate and the associated hopping integrals are weakly orbital dependent $t^{ab}_{ij}\simeq t_{ij}$, 
then the contribution of these two orbitals in $H^{\prime}_{\chi}$ can be simplified as
\begin{equation}
H^{\prime}_{\chi} = - \sum_{(ij)} 2 V_{ij} \left( \widetilde{\boldsymbol{S}}^{\dag}_{ij} 
\cdot \widetilde{\boldsymbol{S}}_{ij} + \frac{1}{4} \widetilde{\chi}^{\dag}_{ij} 
\widetilde{\chi}_{ij} \right) , \label{eqn7.2}
\end{equation}
where $V_{ij} = \frac{2\left( t_{ij}\right)^2}{U}$, and the pseudo-spin operator 
$\widetilde{\boldsymbol{S}}_{ij}$ and the orbital-relevant bond charge operator 
$\widetilde{\chi}_{ij}$ are defined as 
\begin{equation}
\widetilde{\boldsymbol{S}}_{ij} = \sum_{ab\sigma} d^{\dag}_{jb \sigma}\left(
\frac{\boldsymbol{\tau}}{2}\right)_{ba} d_{ia\sigma}, \quad 
\widetilde{\chi}_{ij} = \sum_{a\sigma} d^{\dag}_{ja \sigma} d_{ia\sigma} .  \label{eqn7.3}
\end{equation}
$\boldsymbol{\tau}$ are the Pauli matrices defined in the paired degenerate singly-occupied orbital space. 
The orbital-relevant nematic operators we proposed for FeSCs in Eq. (\ref{eqn2})~\cite{SuLi2013,SuLi2014JPCM} 
are the bond charge operators $\chi_{ij,ab}$ with rotational-symmetry-breaking local form factors. 
Similar to the one-orbital Hubbard model,  the virtual hopping processes also give rise to
a super-exchange interaction in the particle-hole spin channel, 
\begin{equation}
H^{\prime}_{s} = \sum_{(ij) ab} 2V^{ab}_{ij} \left( \boldsymbol{S}_{ia} 
\cdot \boldsymbol{S}_{jb} +\frac{1}{4} n_{ia} n_{jb} \right), \label{eqn8}
\end{equation} 
where $\boldsymbol{S}_{ia}$ and $n_{ia}$ are the on-site spin and charge operators, respectively. 
In the particle-particle channel, $H^{\prime}_{eff}$ has a pairing form as 
\begin{eqnarray}
H^{\prime}_{\Delta} &=& \sum_{(ij) ab} V^{ab}_{ij} 
(\Delta_{ij,ab}^{(s,1)\dag} \Delta^{(s,1)}_{ij,ab} 
+ \Delta_{ij,ab}^{(s,-1)\dag} \Delta^{(s,-1)}_{ij,ab}
\nonumber \\
&+& \frac{1}{2} \Delta_{ij,ab}^{(s,0)\dag} \Delta^{(s,0)}_{ij,ab} 
- \frac{1}{2} \Delta_{ij,ab}^{(a)\dag} \Delta^{(a)}_{ij,ab} ) , \label{eqn9}
\end{eqnarray} 
where the pairing order operators are defined as 
$\Delta^{(s,1)}_{ij,ab} = d_{jb\uparrow} d_{ia\uparrow}$, $\Delta^{(s,-1)}_{ij,ab} 
= d_{jb\downarrow} d_{ia\downarrow}$, and $\Delta^{(s,0)/(a)}_{ij,ab} = 
(d_{jb\uparrow} d_{ia\downarrow} \pm d_{jb\downarrow} d_{ia\uparrow})$.

Some remarks on our approximate treatment of the multi-orbital Hubbard interactions should 
be noted. Firstly, there are other virtual hopping processes in second-order perturbations. 
One special class is described by
$\widetilde{H}^{\prime}_{eff} = - \sum_{ (ij), a\not=b,\sigma_i} 
V^{ab}_{ij} P d^{\dag}_{ib\sigma_2} d_{ja\sigma_2} d^{\dag}_{jb\sigma_1} d_{ia\sigma_1} P $. 
Since each term of $\widetilde{H}^{\prime}_{eff}$ is not self-Hermitian, it describes 
higher-order perturbations compared to $H^{\prime}_{eff}$~\cite{AndersonSuperExchange1959}. 
For examples, $\widetilde{H}^{\prime}_{eff}$ has a new form in the spin channel as
$\widetilde{H}^{\prime}_{s} = \sum_{(ij), a\not=b} 2V^{ab}_{ij} 
\left( \boldsymbol{S}_{i,ab} \cdot \boldsymbol{S}_{j,ab} +\frac{1}{4} n_{i,ab} n_{j,ab} \right)$, 
where $\boldsymbol{S}_{i,ab}=\sum_{\sigma_1 \sigma_2} d^{\dag}_{ib\sigma_1}
\left(\frac{\boldsymbol{\sigma}}{2} \right)_{\sigma_1 \sigma_2} d_{ia\sigma_2}$
and $n_{i,ab} =\sum_{\sigma} d^{\dag}_{ib\sigma} d_{ia\sigma}$ which
describe the inter-orbital spin and charge physics. 
In the pairing channel, $\widetilde{H}^{\prime}_{eff}$ can be reexpressed as 
$
\widetilde{H}^{\prime}_{\Delta} = \sum_{(ij), a\not=b} V^{ab}_{ij} 
(\Delta_{ij,b}^{(s,1)\dag} \Delta^{(s,1)}_{ij,a} + \Delta_{ij,b}^{(s,-1)\dag} \Delta^{(s,-1)}_{ij,a}
+ \frac{1}{2} \Delta_{ij,b}^{(s,0)\dag} \Delta^{(s,0)}_{ij,a} 
- \frac{1}{2} \Delta_{ij,b}^{(a)\dag} \Delta^{(a)}_{ij,a} ) , 
$
where the pairing operators are defined as 
$\Delta^{(s,1)}_{ij,a} = d_{ja\uparrow} d_{ia\uparrow}$, $\Delta^{(s,-1)}_{ij,a} 
= d_{ja\downarrow} d_{ia\downarrow}$, and $\Delta^{(s,0)/(a)}_{ij,a} = 
(d_{ja\uparrow} d_{ia\downarrow} \pm d_{ja\downarrow} d_{ia\uparrow})$.
$\widetilde{H}^{\prime}_{\Delta}$ describes the inter-orbital pairing couplings. 
Secondly, we have neglected the exact configurations of the local orbital occupation in 
the effective Hamiltonian $H^{\prime}_{eff}$, where the interaction constants $V^{ab}_{ij}$
only include the contribution of the paired singly-occupied orbitals $(ia,jb)$. 
In the multi-orbital model for FeSCs, the realistic configuration of the local orbital 
occupation will lead to correction to the effective interaction constants $V^{ab}_{ij}$ 
from local correlations of other orbitals. 
Thirdly, we have not included the Huud's coupling in the above study, which has been 
assumed to be much smaller than $U$. The inclusion of the Huud's coupling would lead to 
slightly spin- and orbital-dependent correction to the effective interaction constants 
$V^{ab}_{ij}$~\cite{AndersonSuperExchange1959,KugelKhomskii1973}, which is ignored at 
our first study. It will lead to a Kugel-Khomskii spin-orbital model for FeSCs~\cite{KrugerZaanen2009}, 
which involves too complex spin-orbital physics to be well studied within an intuitive 
phenomenological theory. Finally, it should be noted that the effective Hamiltonian 
$H^{\prime}_{eff}$ mainly focuses on the low-energy physics driven by the large on-site
Hubbard interactions, where the singly-occupied orbitals are mostly involved. 
Since the crystal-field splitting of the five 3d orbitals and the exact local electronic 
configuration of the Fe ions in FeSCs are not definitely defined, the singly-occupied 
orbitals involved in $H^{\prime}_{eff}$ can be extensively defined. They can include the
orbitals which are singly-occupied and the orbitals which have both finite probability 
to be singly-occupied and finite probability to be doubly-occupied. 
\footnote{A properly realistic phenomenological model for FeSCs can be defined by 
$H=H_t + H_U + H^{\prime}_{eff}$, where $H_U$ are the multi-orbital on-site Hubbard 
interactions including the Huud's coupling, and $H^{\prime}_{eff}$ does not involve 
the Gutzwiller projector $P$. This is one extended multi-orbital model of the $t-U-J$ 
model for the Gossamer superconductivity in the cuprate superconductors~\cite{Laughlin2002,GanZhangPRB2005}.}  

$H^{\prime}_{\chi}$, $H^{\prime}_s$ and $H^{\prime}_{\Delta}$ from the virtual hopping 
processes in the multi-orbital Hubbard model describe different physics in FeSCs in the 
respective particle-hole charge, particle-hole spin and particle-particle channels. Thus 
the bond nematic charge order, the spin magnetic order and the pairing superconducting order 
are intrinsically entangled and can come from the same microscopic virtual hopping processes 
in reduction of the Hubbard energy. 


In the above study, we show that the reduction of the Hubbard energy by virtual hopping processes
can be a driving force of the orbital-relevant nematic charge order. The same virtual hopping 
processes can also lead to a spin $J_1-J_2$ model for FeSCs, which involves an Ising spin nematic 
order~\cite{FangHu2008,CKXu,HuXu}. Thus the orbital-relevant nematic charge order we have 
proposed~\cite{SuLi2013,SuLi2014JPCM} and the Ising spin nematic order come from the same microscopic 
virtual hopping processes in reduction of the Hubbard energy. However, this driving force of 
the orbital-relevant nematic charge order is very different to that of the itinerant magnetic 
nematic order~\cite{Fernandes2012,Fernandes2014}, the latter of which is assumed to stem from 
high-order magnetic fluctuations of the itinerant electrons. One challenge in the scenario of 
the itinerant magnetic nematicity is how to produce a strongly momentum-dependent electronic 
band deformation with d-wave symmetry. The Pomeranchuk nematic orders in the functional and 
parquet renormalization group theories come from the instability of the itinerant electrons
in particle-hole charge channels with zero momentum of center of mass~\cite{Zhai2009,ChubukovPRX2016,
ChubukovPRB942016,ChubukovPRB952017}. The reduction of the nearest-neighbour Hubbard interactions 
as a driving force~\cite{WangZQPRB2016,HuArXiv2016} is different to the one in our theory, 
as the latter of which comes from the reduction of the on-site Hubbard energy by virtual 
hopping processes and can unify the electronic nematicity, the spin magnetism and the pairing 
superconductivity.  

Although the orbital-relevant nematic order we have proposed is one extended orbital-relevant Pomeranchuk 
nematic order, its driving force comes from a strong correlation mechanism. There are two steps 
in energy gain in the formation of the orbital-relevant nematic charge order we proposed. One is 
the reduction of the Hubbard energy in strong correlation processes, and the other is a further 
Fermi-surface instability. Only a pure Fermi-surface instability in weak-coupling scenario drives 
the Pomeranchuk nematic order~\cite{Pomeranchuk1959}. The associated nematic phase transition in 
our theory is one from a normal Gutzwiller projected Fermi liquid to an ordered Gutzwiller projected 
nematic liquid. The nature of this nematic phase transition and the possible difference to the 
Pomeranchuk nematic phase transition~\cite{FradkinNematicNonFL2001,LawlerNonFLnematic2006,
DellAnnaNematicNonFL2006,WuCJNematicFL2007} is one interesting issue for future study. 

As a summary, we have firstly presented all possible electronic nematic operators for FeSCs from 
the Landau's principle of symmetry breaking. We then have presented a driving mechanism that the 
reduction of the Hubbard energy by virtual hopping processes can drive the electrons into a charge 
nematic state. The same mechanism is also the driving force for the spin magnetic state and the 
pairing superconducting state. Following these results, we propose that the bond orbital-relevant 
nematic order in the particle-hole charge channel is the driving nematic order in FeSCs with the 
reduction of the Hubbard energy as the driving force. Moreover, since the electronic nematic order, 
the spin magnetic order and the pairing superconducting order are universal characters of many 
unconventional superconductors~\cite{Fradkin,FisherShen}, they should all stem from one universal 
microscopic mechanism. We propose that the reduction of the Hubbard energy by virtual hopping processes 
is the driving mechanism, and the electronic nematicity, the spin magnetism and the pairing 
superconductivity in unconventional superconductors can then be unified within this mechanism.  

{\it Acknowledgements}
This work was supported by the Natural Science Foundation of China (Grant No. 11674391 ) 
and Research Fund of Renmin University of China. It was also supported partially by 
Shandong Province Natural Science Fund. 



\providecommand{\newblock}{}


\end{document}